\documentclass[%
 reprint,
 amsmath,amssymb,
 aps,
pra,
twocolumn,
]{revtex4-2}

\usepackage{graphicx}
\usepackage{dcolumn}
\usepackage{bm}
\usepackage{bbold}
\usepackage{physics}
\usepackage[colorlinks=true,linkcolor=blue,citecolor=blue,urlcolor=blue]{hyperref}

\begin{document}

\title{Chiral relations for entanglement in quantum self-phase modulation}

\author{M. Gil de Oliveira$^{1,2}$, A. Bramati$^3$, A. Z. Khoury$^1$}
\affiliation{1- Instituto de F\'{i}sica, Universidade Federal Fluminense, 24210-346 Niter\'{o}i, RJ, Brazil}
\affiliation{2 - Quantum Industrial Innovation, Competence Center in Quantum Technologies, Universidade
SENAI-CIMATEC, 40650-010, Salvador, BA, Brazil}
\affiliation{3- Laboratoire Kastler Brossel, Paris, France}

\date{\today}

\begin{abstract}
We investigate entanglement in the radial mode structure generated by self-phase modulation of a coherent beam carrying orbital angular momentum (OAM). We derive a chiral selection rule for the quantum correlations: at specified detection waists, the OAM of the pump and of the detected mode determine a finite set of radial modes whose entanglement can be certified at sufficiently short propagation distances. Finite-distance calculations show that this set continues to organize the certification pattern. Optimized covariance witnesses reveal entanglement involving modes outside that set, with smaller violation margins in the configurations investigated, and certify genuine multipartite entanglement in reduced sets of detection modes. These results identify radial--angular coupling as a means of controlling which spatial mode combinations admit entanglement certification.
\end{abstract}

\maketitle


\section{Introduction}
\label{sec:intro}

The quest for quantum devices based on optical resources is an intense and active field of research, which has already delivered important tools for quantum communication, computing and metrology. This quest has been boosted considerably by the possibility of shaping light in its different degrees of freedom, such as time envelope, spectral and spatial modes \cite{Wilner2015,Fickler2018}. Both discrete and continuous variable quantum information processing can benefit from these developments \cite{Coutinho2009,Gao2014,Gao2018}. Nonlinear optical processes are the common tool for generating entangled quantum states, as they allow the energy transfer between different optical modes, constrained by different conditions such as energy conservation and phase matching, for example \cite{boyd2020nonlinear,Sonja2020,Sonja2021,Sonja2025,Sonja2026}. These conditions constrain the mode couplings through which quantum correlations and entanglement can develop \cite{Jing2019,Jing2020,Jing-Treps-Fabre2020,Jing2021}. Energy and momentum conservation are behind the generation of time-frequency and position-momentum entanglement, respectively. These conditions rule nonlinear optical processes such as three- and four-wave mixing, second harmonic generation and self-phase modulation \cite{Pereira2017,Buono2020,Braian2021-Nanoph,Tabosa1999,Tabosa2003,Tabosa2015,Kaiser2018,Tabosa2023,Motta2022,Motta2023,Motta2024}.

Entanglement provides a resource for quantum communication protocols involving two or more parties~\cite{vanloock2003detecting}. In a spatially structured optical field, orthogonal modes provide distinct subsystems whose quadratures can be addressed by mode-selective detection. Controlling how a nonlinear interaction distributes entanglement among these modes is therefore a step toward preparing spatial resources for quantum information. This requires identifying both the modes involved and the form of entanglement that can be certified: correlations between pairs, entanglement across groups of modes, and genuine multipartite entanglement are different statements about the generated state~\cite{teh2014criteria}.

The possibility of shaping the phase and polarization distributions with spatial light modulators and q-plates has become a powerful tool for controlled generation of customized entangled states both in linear and nonlinear optics. These developments brought the orbital angular momentum (OAM) of light to the toolbox of quantum information technology. Its role in nonlinear processes has also been unveiled by a plethora of scientific papers devoted to the constraints imposed by OAM and vortex formation in nonlinear wave mixing.
In polariton superfluids, the interplay between phase, spin and flow gives rise to a rich variety of spatial structures, including quantized vortices, spin textures and vortex flows \cite{Bramati2016,Bramati2016-2,Bramati2019,Bramati2020,Guerrero2025}. A recent work has further demonstrated the spectroscopy of collective excitations associated with stationary, multiply quantized vortices, opening new possibilities for exploring vortex dynamics in driven-dissipative quantum fluids of light \cite{Guerrero2025}.

A less exploited degree of freedom involves the radial structure of paraxial modes \cite{Pereira2017,Buono2020,Zhi-Han2020,Sonja2021,AOliveira2021,Kotlyar2022}. The Laguerre-Gaussian (LG) solutions of the paraxial wave equation are the natural basis for describing OAM and the radial structure, which are the suitable labels for their identification. These modes are characterized by two integer numbers, namely their topological charge and radial index. The topological charge corresponds to the azimuthal variation of the beam phase around the propagation direction and the radial index determines the number of intensity rings.
While the OAM conservation selection rule has been demonstrated in different quantum optical processes, the constraints related to the radial structure still deserve careful investigation.

Self-phase modulation is a particularly interesting phenomenon, where the intensity of an optical beam alters the effective refractive index of a nonlinear medium, thus affecting the beam's phase evolution along propagation through the medium \cite{Cerda2010}. This effect can also be observed in Bose-Einstein condensates and atomic vapors \cite{BEC-RMP-1999,Mendonca2005,GLORIEUX2025}.
In a previous work we investigated the formation of a radial structure in self-phase modulation driven by an OAM beam and determined the radial spectrum generated in terms of the topological charge of the input beam. This established an interesting radial-angular coupling promoted by the radial selection rules imposed on the nonlinear process \cite{Oliveira2023}.

In the present work we derive a radial--angular selection rule for entanglement certification in self-phase modulation. For a coherent LG pump and specified detection waists, we show that, for sufficiently small propagation distances, covariance-based entanglement witnesses certify pairwise entanglement for a finite set of radial modes. Furthermore, the size of the set is determined by the magnitudes and relative sign of the pump and reference-mode OAM. For larger propagation distances, this set continues to distinguish larger certification margins from the weaker violations involving additional radial modes.

Section~\ref{sec:bosonfield} introduces the spatial mode operators and covariance matrix. Section~\ref{sec:entanglement} introduces the two-mode Duan criterion and shows how pairwise tests and collective covariance witnesses can certify multipartite entanglement. Section~\ref{sec:Kerr} derives the first-order correlations from the full fluctuation equation, and Sec.~\ref{sec:selection rules} establishes their chiral selection rule and its consequences for certification. Section~\ref{sec:finite_distance} examines the finite-distance behavior and genuine multipartite entanglement of reduced sets of detection modes. Section~\ref{sec:experimental_implementation} describes how to measure the mode covariances and distribute the correlated modes to separate parties. Section~\ref{sec:conclusion} summarizes the results.

\section{Spatially structured bosonic field}
\label{sec:bosonfield}

We first introduce the spatial mode operators and correlations used to describe the field. This formulation identifies the optical subsystems and provides the covariance matrix on which the entanglement criteria of Sec.~\ref{sec:entanglement} act.

We start by considering the normalized positive frequency component of the quantized electromagnetic field
$\mathcal{E}(\mathbf{r})$ satisfying the commutation relations
\begin{equation}
    \label{eq:commutation_relations}
    \begin{aligned}
        \comm{\mathcal{E}(\mathbf{r}_1)}{\mathcal{E}^\dagger(\mathbf{r}_2)}
        &= \delta^{(2)}(\mathbf{r}_1 - \mathbf{r}_2), \\
        \comm{\mathcal{E}(\mathbf{r}_1)}{\mathcal{E}(\mathbf{r}_2)}
        &= \comm{\mathcal{E}^\dagger(\mathbf{r}_1)}{\mathcal{E}^\dagger(\mathbf{r}_2)} = 0.
    \end{aligned}
\end{equation}
In the above equation and in what follows, $\mathbf{r} = (x, y)$ is the transverse coordinate.
One can introduce ladder operators $a(v), a^\dagger(v)$ associated with the spatial mode $v$ by the expression
\begin{equation}
    a(v) = \int_{\mathbb{R}^2} v^*(\mathbf{r}) \mathcal{E}(\mathbf{r}) d\mathbf{r}.
\end{equation}
We then have
\begin{eqnarray}
    \comm{a(v_1)}{a^\dagger(v_2)} &=& (v_1;v_2) \equiv\int_{\mathbb{R}^2} v_1^*(\mathbf{r}) v_2(\mathbf{r}) d\mathbf{r}\,,
    \nonumber\\
    \comm{a(v_1)}{a(v_2)} &=& \comm{a^\dagger(v_1)}{a^\dagger(v_2)} = 0\,.
\end{eqnarray}
For orthonormal profiles, $(v_j;v_k)=\delta_{jk}$, these are independent bosonic mode operators. Each mode has its own Fock space, and a set of such modes defines a tensor product of subsystems. Orthogonality does not require spatially separated intensity profiles: distinct modes can overlap within the same beam. The operator construction also applies to nonorthogonal or unnormalized detection functions, but these do not by themselves define independent parties for an entanglement test. All mode parties considered below are defined by orthonormal profiles.

Given a classical field distribution $\alpha(\mathbf{r})\,$ normalized in such a way that $\abs{\alpha(\mathbf{r})}^2$ represents the photon number per unit area, we may build a coherent state through the expression
\begin{eqnarray}
    \label{eq:initial-state}
    \ket{\alpha} = \exp\left[a^\dagger(\alpha) - a(\alpha)\right] \ket{0}\,,
\end{eqnarray}
where $\ket{0}$ is the vacuum state. This state is an eigenvector of the field operator,
\begin{equation}
    \mathcal{E}(\mathbf{r}) \ket{\alpha} = \alpha(\mathbf{r}) \ket{\alpha}\;.
\end{equation}
This is the type of state that will be used as input on the nonlinear medium.

The field quadratures are defined by
\begin{equation}
    X(v) = \frac{a(v) + a^\dagger(v)}{\sqrt{2}}\,.
\end{equation}
Here, $v$ is the corresponding field mode. The complementary quadrature is simply defined by shifting $v$ by a $\pi/2$ phase: $P(v) = X(iv)$.

The covariance function is given by
\begin{equation}
    \label{eq: covariance}
    \Gamma(v_1, v_2) = \expval{\acomm{\Delta X(v_1)}{\Delta X(v_2)}},
\end{equation}
where $\Delta X(v) = X(v) - \expval{X(v)}$ and $\acomm{\bullet}{\bullet}$ denotes the anticommutator. The calculation of this observable is simplified if we express it in terms of the field $\mathcal{E}$. For this, we introduce the normal and anomalous correlations, given respectively by
\begin{equation}
    \label{eq:FHdefinitions}
    H(\mathbf{r}_1, \mathbf{r}_2)
        = \expval{\mathcal{E}^\dagger(\mathbf{r}_2) \mathcal{E}(\mathbf{r}_1)}
        - \expval{\mathcal{E}^\dagger(\mathbf{r}_2)}
           \expval{\mathcal{E}(\mathbf{r}_1)} \,,
\end{equation}
\begin{equation}
    F(\mathbf{r}_1, \mathbf{r}_2)
        = \expval{\mathcal{E}(\mathbf{r}_1) \mathcal{E}(\mathbf{r}_2)}
           - \expval{\mathcal{E}(\mathbf{r}_1)}\expval{\mathcal{E}(\mathbf{r}_2)}.
\end{equation}

We may then write
\begin{equation}
    \label{eq:quadrature_correlation}
    \begin{aligned}
        \Gamma(v_1, v_2) = \Re\left[(v_1; v_2) + 2f(v_1, v_2) + 2h(v_1, v_2)\right],
    \end{aligned}
\end{equation}
where
\begin{equation}
\label{eq:fhdefinitions}
    \begin{aligned}
        h(v_1, v_2) &= \int v_1^*(\mathbf{r}_1) v_2(\mathbf{r}_2)
           H(\mathbf{r}_1, \mathbf{r}_2)\,d\mathbf{r}_1\,d\mathbf{r}_2 \\ &= \expval{a^\dagger(v_2)a(v_1)} - \expval{a(v_2)}^* \expval{a(v_1)},
    \end{aligned}
\end{equation}
and
\begin{equation}
    \begin{aligned}
        f(v_1, v_2) &= \int v_1^*(\mathbf{r}_2) v_2^*(\mathbf{r}_1)
           F(\mathbf{r}_1, \mathbf{r}_2)\,d\mathbf{r}_1\,d\mathbf{r}_2 \\ &= \expval{a(v_1)a(v_2)} - \expval{a(v_1)} \expval{a(v_2)}.
    \end{aligned}
\end{equation}

We further note the following properties, which simplify the consideration of linear combination of modes:
\begin{equation}
    \begin{aligned}
        f(v_1, v_2 + \lambda v_3) &= f(v_1, v_2) + \lambda^* f(v_1, v_3); \\
        f(v_1, v_2) &= f(v_2, v_1); \\
        h(v_1, v_2 + \lambda v_3) &= h(v_1, v_2) + \lambda h(v_1, v_3); \\
        h(v_1, v_2) &= h(v_2, v_1)^*.
    \end{aligned}
\end{equation}
These are a reflection of the fact that $a(v)$ is antilinear while $a^\dagger(v)$ is linear.

For $N$ orthonormal modes, define the quadrature vector and covariance matrix by
\begin{equation}
    \begin{aligned}
        \mathbf R&=\bigl(X(v_1),P(v_1),\ldots,X(v_N),P(v_N)\bigr),\\
        \gamma_{jk}&=\expval{\acomm{\Delta R_j}{\Delta R_k}}.
    \end{aligned}
    \label{eq:multimode_covariance}
\end{equation}
Every entry follows from Eq.~\eqref{eq:quadrature_correlation}, using $P(v)=X(iv)$. With our normalization, $[X(v_j),P(v_k)]=i\delta_{jk}$, the vacuum variance of each quadrature is $1/2$, and the vacuum covariance matrix is $\mathbb 1_{2N}$.

\section{Entanglement and covariance-based certification}
\label{sec:entanglement}

In this section we investigate under what conditions do the quadrature covariances of two orthogonal modes certify entanglement, and what can such pairwise tests tell us about a larger set of modes. This will set the toolbox for the entanglement analysis in the mode structure generated by self-phase modulation.

\subsection{Pairwise Entanglement}
\label{sec:duan}

Let $\rho$ denote the reduced state of two orthonormal modes $v_1$ and $v_2$, obtained by tracing out all other field modes. The state $\rho$ is separable if it can be decomposed as
\begin{equation}
    \rho=\sum_\lambda p_\lambda
        \rho_1^{(\lambda)}\otimes\rho_2^{(\lambda)},
    \qquad p_\lambda\ge0,\quad \sum_\lambda p_\lambda=1,
\end{equation}
where $\rho_j^{(\lambda)}$ is a state in the Hilbert space of the single mode $v_j$. Otherwise, it is said to be entangled. We can test this by employing the Duan criterion~\cite{duan2000inseparability, PhysRevLett.84.2726}. In the normalization of Sec.~\ref{sec:bosonfield} it certifies entanglement when
\begin{equation}
    \begin{aligned}
        D &= \frac{1}{2}\expval{\left[\Delta X(v_1)+\Delta X(v_2)\right]^2}\\
          &\quad+\frac{1}{2}\expval{\left[\Delta P(v_1)-\Delta P(v_2)\right]^2}<1.
    \end{aligned}
    \label{eq:duan_criterion}
\end{equation}
Expressing the variances in terms of the mode correlations in Eq.~\eqref{eq:quadrature_correlation} gives
\begin{equation}
    D=1+h(v_1,v_1)+h(v_2,v_2)+2\Re f(v_1,v_2).
\end{equation}
Rotating the quadrature phase of $v_1$ leaves both normal correlations $h$ unchanged. We choose this phase so that the anomalous correlation $f$ contributes its most negative real part, obtaining the phase-optimized Duan witness
\begin{equation}
    \label{eq:optimized_duan}
    D_{\mathrm{opt}}=1+h(v_1,v_1)+h(v_2,v_2)-2\abs{f(v_1,v_2)}.
\end{equation}
For the coherent input, both $h$ and $f$ vanish. As the field propagates through the Kerr medium, a pair violates the Duan bound whenever $2|f(v_1,v_2)|$ exceeds the sum of its two normal correlations. This is the condition we evaluate at short distances in Sec.~\ref{sec:Kerr}.

\subsection{Multimode Entanglement}
\label{sec:multimode_certification}

Now consider $N>2$ orthonormal modes $v_1,\ldots,v_N$, and let $\rho$ denote their joint reduced state. A bipartition $\mathcal S|\mathcal T$ divides these $N$ modes into two nonempty groups. As in the two-mode case, $\rho$ is separable across this bipartition if it can be written as
\begin{equation}
    \rho=\sum_\lambda p_\lambda
        \rho_\mathcal{S}^{(\lambda)}\otimes\rho_\mathcal{T}^{(\lambda)},
    \qquad p_\lambda\ge0,\quad \sum_\lambda p_\lambda=1,
    \label{eq:cut_separable}
\end{equation}
where $\rho_\mathcal{S}^{(\lambda)}$ and $\rho_\mathcal{T}^{(\lambda)}$ are states of the two parties. Otherwise it is entangled across this bipartition.

Suppose two modes, one in $\mathcal S$ and one in $\mathcal T$, violate the Duan bound. If $\rho$ were separable across this cut, tracing out the other $N-2$ modes in Eq.~\eqref{eq:cut_separable} would leave the pair in a mixture of product states, contradicting the violation. Thus an entangled pair certifies entanglement across every cut that separates its two modes. If every bipartition separates at least one entangled pair, $\rho$ is entangled across every bipartition and is therefore \emph{fully inseparable}~\cite{vanloock2003detecting}.

Even when the complete field is pure, the reduced state of these $N$ modes can be mixed. A preparation could produce states separable across one cut on some runs and across a different cut on others. The resulting mixture, called biseparable, has the form
\begin{equation}
    \rho_{\mathrm{bisep}}=\sum_\lambda p_\lambda
        \rho_{\mathcal{S}_\lambda|\mathcal{T}_\lambda},
    \label{eq:biseparable_mixture}
\end{equation}
where each component admits the decomposition in Eq.~\eqref{eq:cut_separable} for its own cut $\mathcal S_\lambda|\mathcal T_\lambda$. Such a mixture can be entangled across every fixed cut. Certifying genuine multipartite entanglement (GME) therefore requires excluding all decompositions of the form~\eqref{eq:biseparable_mixture}~\cite{teh2014criteria}. For a pure state, full inseparability already suffices.

\subsection{Optimized covariance witnesses}
\label{sec:covariance_witnesses}

To extend the test to collective quadrature correlations, we first write the Duan sum in terms of the two-mode covariance matrix $\gamma_{12}$ of $v_1$ and $v_2$. In the quadrature ordering of Eq.~\eqref{eq:multimode_covariance}, $D=\operatorname{Tr}(\mathcal Z_D\gamma_{12})$ with
\begin{equation}
    \mathcal Z_D=\frac14\begin{pmatrix}
        1&0&1&0\\
        0&1&0&-1\\
        1&0&1&0\\
        0&-1&0&1
    \end{pmatrix}.
    \label{eq:duan_witness}
\end{equation}
For $N$ modes with covariance matrix $\gamma$, the coefficients in a real, symmetric, positive semidefinite $2N\times 2N$ matrix $\mathcal Z$ specify a weighted sum of covariances. To use this sum as a witness, we choose a separability class we wish to exclude and require
\begin{equation}
    \operatorname{Tr}(\mathcal Z\gamma_{\mathrm{sep}})\ge1,
    \label{eq:witness_bound}
\end{equation}
for every covariance matrix $\gamma_{\mathrm{sep}}$ in that class~\cite{hyllus2006optimal}. A measured or calculated covariance matrix $\gamma$ satisfying
\begin{equation}
    \operatorname{Tr}(\mathcal Z\gamma)<1,
    \label{eq:witness_violation}
\end{equation}
then excludes the chosen class. A value at or above one is inconclusive. The margin below one gives the size of the witness violation and does not quantify the entanglement.

In the finite-distance calculations, we minimize this weighted sum for the reduced covariance matrix of the modes under consideration using the semidefinite programs of Hyllus and Eisert~\cite{hyllus2006optimal}. To test a fixed bipartition, the FullyWit routine imposes Eq.~\eqref{eq:witness_bound} on states separable across that cut. This may allow correlations among all $N$ modes to certify the cut even when no individual pair yields a Duan violation. To test for GME, the MultiWit routine imposes the bound on the entire biseparable class. A single violation then excludes every mixture in Eq.~\eqref{eq:biseparable_mixture}.

\section{Quantum correlations in the short-range Kerr dynamics}

\label{sec:Kerr}

We consider a monochromatic field propagating through a $\chi^{(3)}$ nonlinear medium, with a coherent spatial mode as input. The nonlinear interaction produces both a change in the mean field and correlations between mode fluctuations. We now calculate these correlations to first order in propagation distance and apply the criteria of Sec.~\ref{sec:entanglement}.

We will assume that the field is propagating along the $z$ direction,
and satisfies the Heisenberg equation of motion
\begin{equation}
    \label{eq:heisenberg_equation}
    2ik \partial_z \mathcal{E}(\mathbf{r}, z) + \nabla^2 \mathcal{E}(\mathbf{r}, z) = g\, \mathcal{E}^\dagger(\mathbf{r}, z) \mathcal{E}^2 (\mathbf{r}, z)\,,
\end{equation}
where $g$ is the Kerr nonlinear coupling constant and $\nabla^2$ denotes the transverse Laplacian. The short-distance expansion below retains the full fluctuation equation and yields the exact coefficient linear in distance. In Sec.~\ref{sec:finite_distance}, we adopt a different approximation: diffraction-free, linearized fluctuation dynamics with the full distance dependence retained. This allows us to investigate how the first-order certification pattern extends to finite propagation distances.

To start our derivation, let $\alpha_c(\mathbf{r}, z)$ denote the solution of the classical equations of motion
\begin{equation}
    2ik \partial_z \alpha_c + \nabla^2 \alpha_c = g\, \abs{\alpha_c}^2 \alpha_c,
\end{equation}
under the initial condition $\alpha_c(\mathbf{r}, 0) = \alpha(\mathbf{r})$, where $\ket{\alpha}$ is the input coherent state. Let us also define the fluctuation field
\begin{equation}
    \eta(\mathbf{r}, z) = \mathcal{E}(\mathbf{r}, z) - \alpha_c(\mathbf{r}, z).
\end{equation}

We observe that our initial state has vacuum-like properties with respect to $\eta(\mathbf{r}, 0)$. For example, it is annihilated by $\eta(\mathbf{r}, 0)$:
\begin{equation}
    \eta(\mathbf{r}, 0) \ket{\alpha} = 0.
\end{equation}
Furthermore, any normally ordered monomial containing at least one fluctuation operator has zero expectation value at $z=0$.

The correlation functions, which are our observables of interest, can also be compactly written in terms of $\eta$:
\begin{equation}
\begin{aligned}
    F(\mathbf r_1,\mathbf r_2;z)&=\langle\eta(\mathbf{r}_1, z)\eta(\mathbf{r}_2, z)\rangle-\langle\eta(\mathbf{r}_1, z)\rangle\langle\eta(\mathbf{r}_2, z)\rangle,\\
    H(\mathbf r_1,\mathbf r_2;z)&=\langle\eta^\dagger(\mathbf{r}_2, z)\eta(\mathbf{r}_1, z)\rangle-\langle\eta(\mathbf{r}_1, z)\rangle\langle\eta^\dagger(\mathbf{r}_2, z)\rangle.
\end{aligned}
\end{equation}

Next, we write the equation of motion
\begin{equation}
    \label{eq:bg_full_fluctuation}
    2ik\partial_z\eta+\nabla^2\eta=g\left(2|\alpha_c|^2\eta+\alpha_c^2\eta^\dagger+\Omega\right),
\end{equation}
where the nonlinear fluctuation terms are
\begin{equation}
    \Omega=\alpha_c^*\eta^2
            +2\alpha_c\eta^\dagger\eta
            +\eta^\dagger\eta^2.
    \label{eq:bg_remainder}
\end{equation}

We start by considering an approximation where only terms up to $\mathcal{O}(z)$ are retained. By definition, we already have $\expval{\eta(\mathbf{r}, 0)} = 0$. Furthermore, as the equation of motion \eqref{eq:bg_full_fluctuation} contains only normally ordered terms, we also get $\partial_z \expval{\eta(\mathbf{r}, 0)} = 0$, such that
\begin{equation}
    \expval{\eta(\mathbf{r}, z)} = \mathcal{O}(z^2),
\end{equation}
or, equivalently,
\begin{equation}
    \expval{\mathcal{E}(\mathbf{r}, z)} = \alpha_c(\mathbf{r}, z) + \mathcal{O}(z^2).
    \label{eq:exp_val_field}
\end{equation}

Now we turn to the correlation functions. For compactness, we define $\eta_{j,z} \equiv \eta(\mathbf{r}_j, z)$. In order to calculate the normal correlation, we differentiate the two-operator average:
\begin{equation}
    \left.\partial_z\langle\eta_2^\dagger\eta_1\rangle\right|_0
       =\langle(\partial_z\eta_2^\dagger)_0\eta_{1,0}\rangle
         +\langle\eta_{2,0}^\dagger(\partial_z\eta_1)_0\rangle.
\end{equation}
Both terms vanish: the first average has an annihilation operator on the right, and the second has its conjugate on the left.

Turning to the anomalous case, we have that
\begin{equation}
\begin{aligned}
        \left.\partial_z\langle\eta_{1,z}\eta_{2,z}\rangle\right|_0
       &= \langle(\partial_z\eta_1)_0\eta_{2,0}\rangle
          +\langle\eta_{1, 0}(\partial_z\eta_2)_0\rangle \\&= \langle\eta_{1, 0}(\partial_z\eta_2)_0\rangle.
\end{aligned}
\end{equation}

Isolating $\partial_z \eta$ in the equations of motion \eqref{eq:bg_full_fluctuation} and substituting above, we see that almost all the terms are zero, because they contain $\eta$ to the right. The only exception is the term proportional to $\eta^\dagger$, which gives
\begin{equation}
    \left.\partial_z\langle\eta_1\eta_2\rangle\right|_0
       =-\frac{ig}{2k}\alpha^2(\mathbf r_2)
                     \delta^{(2)}(\mathbf r_1-\mathbf r_2).
\end{equation}

Since the mean-subtraction terms do not contribute at first order, the correlations are
\begin{equation}
\begin{aligned}
    F(\mathbf r_1,\mathbf r_2;z)
      &=-\frac{igz}{2k}\alpha^2(\mathbf r_2)
             \delta^{(2)}(\mathbf r_1-\mathbf r_2)+\mathcal{O}(z^2),\\
    H(\mathbf r_1,\mathbf r_2;z)&=\mathcal{O}(z^2).
\end{aligned}
    \label{eq:bg_first_correlations}
\end{equation}
Using the spatial overlap notation
\begin{equation}
    (v_1,\ldots,v_n;u_1,\ldots,u_n)
       =\int d^2\mathbf r\,\prod_{j=1}^n v_j^*(\mathbf r)u_j(\mathbf r),
\end{equation}
projection onto the modes $v_1$ and $v_2$ gives
\begin{equation}
\begin{aligned}
    f(v_1,v_2)&=-\frac{igz}{2k}(v_1,v_2;\alpha,\alpha)+\mathcal{O}(z^2),\\
    h(v_1,v_2)&=\mathcal{O}(z^2).
\end{aligned}
\end{equation}

To specify the input and introduce a natural transverse scale, we consider coherent Laguerre-Gaussian modes, with
\begin{equation}
    \alpha(\mathbf r)=\norm{\alpha}\,LG_{pl}(\mathbf r;w),
\end{equation}
where the waist-plane modes are
\begin{equation}
    \begin{aligned}
        LG_{pl}(\mathbf{r}; w) &= \frac{(-1)^p\mathcal{N}_{pl}}{w}
        \left(\frac{2r^2}{w^2}\right)^{|l|/2}\\
        &\quad\times L_p^{|l|}\!\left(\frac{2r^2}{w^2}\right)
        e^{-r^2/w^2+il\phi},
    \end{aligned}
\end{equation}
with radial order $p$, topological charge $l$, and normalization constant
\begin{equation}
    \mathcal{N}_{pl}=\sqrt{\frac{2}{\pi}}\sqrt{\frac{p!}{(p+|l|)!}}\;.
\end{equation}
The waist $w$ sets the natural transverse scale. Let us define the Rayleigh range
$z_R=kw^2/2$, the nonlinear interaction length $z_{\mathrm{NL}}=4z_R/(g\norm{\alpha}^2)$,
the dimensionless propagation distance $Z=z/z_{NL}=gz\norm{\alpha}^2/(2kw^2)$,
and the dimensionless input profile $u(\mathbf r)=w\alpha(\mathbf r,0)/\norm{\alpha}$.

Using these definitions, Eq.~\eqref{eq:quadrature_correlation} and the Duan criterion give the first-order results
\begin{equation}
    \begin{aligned}
        f(v_1,v_2)&\approx-iZ(v_1,v_2;u,u),\\
        \Gamma(v_1,v_2)&=\Re(v_1;v_2)+2Z\Im(v_1,v_2;u,u)+\mathcal{O}(Z^2),\\
        D_{\mathrm{opt}}&=1-2\left|Z(v_1,v_2;u,u)\right|+\mathcal{O}(Z^2).
    \end{aligned}
\end{equation}
For an input angular momentum $l_0$, the angular integration requires $l_{v_1}+l_{v_2}=2l_0$ for a linear Duan violation. This OAM condition is necessary for first-order certification; the corresponding radial overlap must also be nonzero.

\section{Chiral selection rule and entanglement certification}
\label{sec:selection rules}

We now specialize to the LG input $\alpha(\mathbf r)=\norm{\alpha}\,LG_{0,l_0}(\mathbf r;w)$ and analyze the selection rules for radial and angular mode coupling.

Our previous work \cite{Oliveira2023} has shown that, in the classical regime, the first-order treatment gives a selection rule for the generation of radial orders up to $\abs{l_0}$. The current quantum treatment clarifies that this was, in fact, a selection rule for the mean field value $\expval{\mathcal{E}}$ given by \eqref{eq:exp_val_field}, whose expression had already appeared in our earlier treatment.

We now derive the radial selection rule for the first-order anomalous correlations and use it to establish pairwise and multimode entanglement at sufficiently small nonzero distances.

\subsection{Radial selection rules and two-mode entanglement}

Let $v_2 = LG_{0,l_2}(\mathbf{r}; w_2)$ be our reference mode, where $w_2$ is free and $l_2$ is only constrained by $l_2 \ne l_0$. Now, define $l_1 = 2l_0-l_2$ and $1/w_1^2 = 2/w^2 + 1 /w_2^2$. We then have that
\begin{eqnarray}
        v_2^*(\mathbf{r})u^2(\mathbf{r})
        &=& \frac{\mathcal{N}_{0l_2}\,\mathcal{N}_{0l_0}^2}{w_2}
        \left(\frac{2r^2}{w_2^2}\right)^{\!\!\!\frac{\abs{l_2}}{2}}\!\!\!
        \left(\frac{2r^2}{w^2}\right)^{\!\!\abs{l_0}} \!\!\!e^{-\frac{r^2}{w_1^2}+il_1\phi}
        \nonumber\\
        &=& \frac{\mathcal{N}_{0l_2}\,\mathcal{N}_{0l_0}^2 \,w_1^{2\abs{l_0}+\abs{l_2}}}
        {w_2^{\abs{l_2}+1}w^{2\abs{l_0}}}
        \left(\frac{2r^2}{w_1^2}\right)^{\!\!\!\frac{\abs{l_1}}{2}}\!\!
        \left(\frac{2r^2}{w_1^2}\right)^{p_{\text{max}}}
        \nonumber\\
        &\times& e^{-\frac{r^2}{w_1^2}+il_1\phi}\;,
\end{eqnarray}
where we have defined
\begin{eqnarray}
    \label{eq:radial_cutoff}
    p_{\text{max}} &=& \abs{l_0}+\frac{\abs{l_2} - \abs{2l_0 - l_2}}{2}
    \nonumber\\
    &=& \begin{cases}
        0, & l_0l_2 \le 0 \\
        \min(2\abs{l_0}, \abs{l_2}), & l_0l_2 > 0
    \end{cases}
    \,.
\end{eqnarray}
Then, by expanding the monomial that is raised to the power of $p_{\text{max}}$ in terms of Laguerre polynomials, we get
\begin{equation}
    v_2^*(\mathbf{r}) u^2(\mathbf{r}) = \sum_{p=0}^{p_{\text{max}}} C_{p}\,LG_{pl_1}(\mathbf{r}; w_1)\;,
\end{equation}
where
\begin{equation}
    \label{eq:radial_coefficients}
    C_p = \frac{p_{\text{max}}!\,\mathcal{N}_{0l_2} \,\mathcal{N}_{0l_0}^2 \,w_1^{2\abs{l_0} + \abs{l_2} + 1}}{\mathcal{N}_{pl_1}w_2^{\abs{l_2}+1}w^{2\abs{l_0}}} \binom{p_{\text{max}} + \abs{l_1}}{p_{\text{max}} - p}\;.
\end{equation}
Therefore, setting $v_1 = LG_{pl_1}(\mathbf{r}; w_1)$, which is orthogonal to $v_2$, and considering the two-mode subsystem formed by $v_1, v_2$, we get an optimal Duan violation of
\begin{equation}
    \label{eq:radial_duan}
    D_{\text{opt}} = \begin{cases}
        1 - 2C_p \abs{Z}+\mathcal{O}(Z^2), & p \le p_{\text{max}} \\
        1+\mathcal{O}(Z^2), & p > p_{\text{max}}
    \end{cases}\;.
\end{equation}
For $0\le p\le p_{\max}$, the coefficients $C_p$ are positive with the chosen LG phase convention. Their linear contribution therefore dominates the $\mathcal{O}(Z^2)$ remainder for sufficiently small nonzero $Z$, establishing entanglement of each corresponding pair. For $p>p_{\max}$, the linear coefficient vanishes: Eq.~\eqref{eq:radial_duan} does not decide whether higher-order correlations produce entanglement. The finite radial support is a property of this first-order coefficient in the specified waist basis.

Throughout the subsequent analysis, we set the reference-mode waist equal to the pump waist, $w_2=w$. The matching condition $1/w_1^2=2/w^2+1/w_2^2$ then fixes the partner-mode waist to $w_1=w/\sqrt{3}$.

\begin{figure*}[!t]
    \centering
    \includegraphics[width=0.82\linewidth]{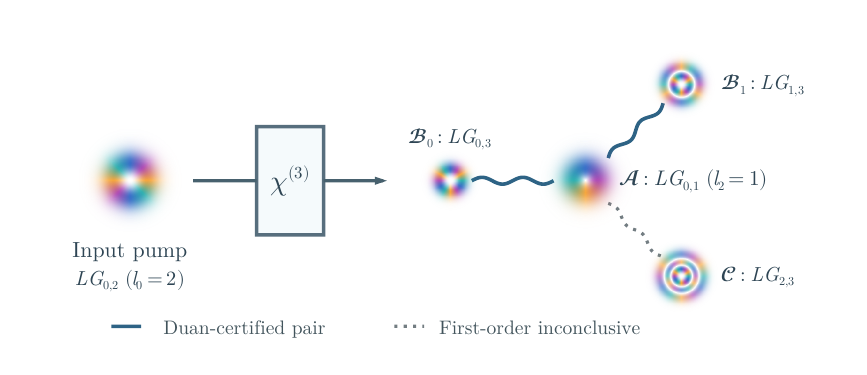}
    \caption{Schematic of radial--angular mode coupling and pairwise certification for an OAM pump with $l_0=2$. The nonlinear medium generates an output field containing the reference mode $\mathcal A=LG_{0,1}(w_2=w)$ and radial partners $\mathcal B_0=LG_{0,3}(w_1)$, $\mathcal B_1=LG_{1,3}(w_1)$, and $\mathcal C=LG_{2,3}(w_1)$. Here $l_2=1$, $l_1=3$, and $p_{\max}=1$. Solid links indicate the first-order Duan-violating pairs $\mathcal A|\mathcal B_p$ for $p=0,1$; the dashed link marks $\mathcal A|\mathcal C$ as inconclusive at first order, not as unentangled. Phase is encoded by hue and intensity by transparency in the mode profiles.}
    \label{fig:selection_schematic}
\end{figure*}

Figure~\ref{fig:selection_schematic} illustrates the mode selection for $l_0=2$ and $l_2=1$, for which $l_1=3$ and $p_{\max}=1$. This is the same input and mode family used in the multimode witness analysis that follows.

\subsection{Consequences for multipartite certification}
\label{sec:pairwise_partitions}

Let $\mathcal A$ denote the reference mode $v_2$, and let $\mathcal B_p$ denote $LG_{p,l_1}(w_1)$ for $p=0,\ldots,p_{\max}$. These are mutually orthonormal mode parties. Since their number is finite, Eq.~\eqref{eq:radial_duan} gives a common small-distance interval in which every reduced pair $\mathcal A|\mathcal B_p$ is entangled. Every nontrivial bipartition of the selected set separates $\mathcal A$ from at least one partner. The crossing-pair criterion of Sec.~\ref{sec:multimode_certification} therefore establishes full inseparability of this set. This applies to its generally mixed reduced state and does not, by itself, certify GME.

For the ideal closed, lossless model, purity of the complete field gives a further consequence. Denote the remaining field modes by $\mathcal R$. For $p_{\max}\ge1$, the grouping
\begin{equation}
    \mathcal A\;|\;\mathcal B_0\;|\;\cdots\;|\;\mathcal B_{p_{\max}-1}\;|\;(\mathcal B_{p_{\max}}\mathcal R),
    \label{eq:main_grouped_parties}
\end{equation}
includes all degrees of freedom and is pure. The same pairs cross every bipartition, so these $p_{\max}+2$ grouped parties are genuinely multipartite entangled. For $p_{\max}=0$, the statement reduces to bipartite entanglement across $\mathcal A|(\mathcal B_0\mathcal R)$. This party count is a guaranteed grouping at the specified detection waists, not an upper bound on the number of entangled field modes. It also does not establish GME of the selected single modes after $\mathcal R$ is discarded.

We next ask how the certification pattern changes at finite distance and whether GME can be certified directly in the reduced selected state.

\section{Entanglement beyond the short-range regime}
\label{sec:finite_distance}

In this section we treat the field dynamics beyond the short-range regime, still within the diffraction-free limit. In other words, we consider $z_{\mathrm{NL}} \lesssim z \ll z_R$. As we will see, the main qualitative aspects of the short-range behavior remain valid.

\subsection{Diffraction-free Bogoliubov solution}

The first-order Duan expression describes the initial generation of correlations but cannot be extrapolated to finite distances: Eq.~\eqref{eq:radial_duan} may become negative for large enough $\abs{Z}$. For finite-distance predictions, we return to the full equation for the same decomposition $\mathcal E=\alpha_c+\eta$ and adopt a different approximation: we neglect diffraction and discard the nonlinear fluctuation terms $\Omega$ in Eq.~\eqref{eq:bg_full_fluctuation}, while retaining the full dependence on propagation distance. This linearization of the fluctuation dynamics is the Bogoliubov approximation, used to describe quantum fluctuations in paraxial propagation through Kerr media~\cite{PhysRevA.92.043802}. Appendix~\ref{app:no_diffraction} solves the diffraction-free Bogoliubov equations without a Taylor expansion in distance. The dimensionless classical field is then
\begin{equation}
    u_Z(\mathbf r)=u(\mathbf r)e^{-iZ|u(\mathbf r)|^2},
\end{equation}
and the mode correlations are
\begin{equation}
\begin{aligned}
    f(v_1,v_2)={}&-iZ(v_1,v_2;u_Z,u_Z)\\
                &-Z^2(v_1,v_2,u_Z;u_Z,u_Z,u_Z),\\
    h(v_1,v_2)={}&Z^2(v_1,u_Z,u_Z;v_2,u_Z,u_Z).
\end{aligned}
\end{equation}
These expressions agree with the preceding first-order result and are exact within the diffraction-free linearized fluctuation equations. Their use at finite distance requires negligible diffraction over the interaction region, which is a reasonable approximation for softly focused beams, as well as a small fluctuation backreaction. The numerical results below use these correlations under those assumptions.

\subsection{Two-mode certification}

\begin{figure*}[!t]
    \centering
    \includegraphics[width=\linewidth]{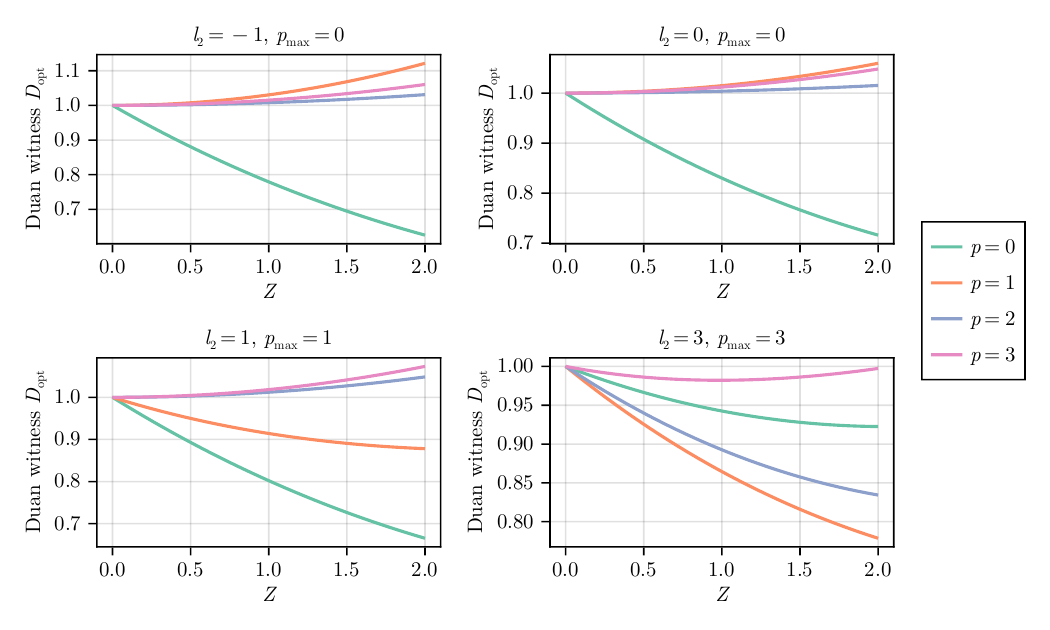}
    \caption{Phase-optimized Duan witness $D_{\mathrm{opt}}$ for the diffraction-free Bogoliubov solution with fixed pump OAM $l_0=2$ and reference-mode OAM $l_2=-1,0,1,3$. Curves correspond to radial indices $p=0,\ldots,3$ of the partner mode, with $l_1=4-l_2$, $w_1=w/\sqrt{3}$, and $w_2=w$. Values below one certify two-mode entanglement.}
    \label{fig:duan}
\end{figure*}

Figure~\ref{fig:duan} compares the phase-optimized Duan witness at fixed $l_0=2$ for $l_2=-1,0,1,3$. The corresponding first-order radial cutoffs are $p_{\max}=0,0,1,3$, respectively. The first two panels illustrate the single radial partner selected for opposite-sign or zero reference-mode OAM, while the latter two illustrate the larger radial support available for equal-sign OAM. In particular, all four plotted radial indices lie within the first-order support for $l_2=3$. The $l_2=1$ panel uses the same input and radial partners as the schematic in Fig.~\ref{fig:selection_schematic} and the multimode witness analysis below.

\subsection{Entanglement across multimode bipartitions}

\begin{figure*}[!t]
    \begin{minipage}[t]{0.48\textwidth}
    \vspace{0pt}
    \centering
    \includegraphics[width=\linewidth]{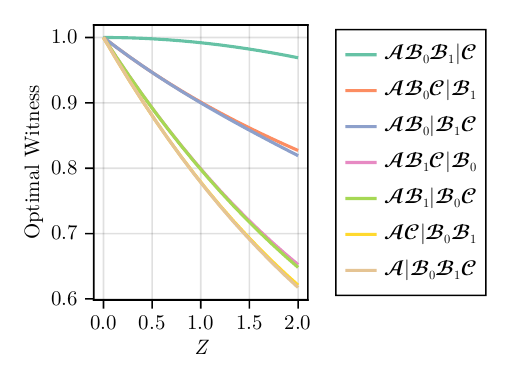}
    \caption{Optimized covariance witnesses for the seven bipartitions of the reduced four-mode state $\mathcal A\mathcal B_0\mathcal B_1\mathcal C$. Values below one certify entanglement across the corresponding cut. Here $l_0 = 2$ and $l_2 = 1$, so that $l_1=2l_0-l_2=3$ and $p_{\max}=\min(2\abs{l_0},\abs{l_2})=1\,$.}
    \label{fig:fully_wit}
    \end{minipage}\hfill
    \begin{minipage}[t]{0.48\textwidth}
    \vspace{0pt}
    \centering
    \includegraphics[width=\linewidth]{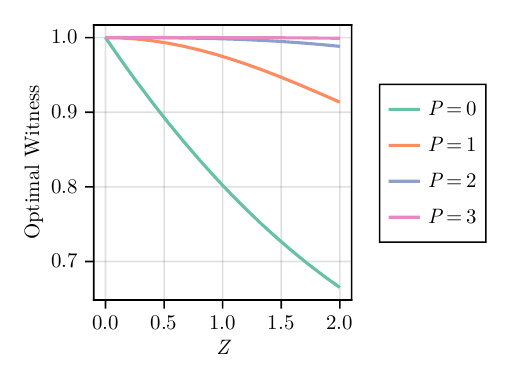}
    \caption{Optimized witnesses for the reduced selected sets containing $\mathcal A$ and radial partners $0\le p\le P$. Values below one certify bipartite entanglement for $P=0$ and genuine multipartite entanglement for $P\ge1$. Here $l_0 = 2$ and $l_2 = 1$, so that $l_1=2l_0-l_2=3$ and $p_{\max}=\min(2\abs{l_0},\abs{l_2})=1\,$.}
    \label{fig:multi_wit}
    \end{minipage}
\end{figure*}

Now we turn to more parties. We set $l_0 = 2$ and $l_2 = 1$, so that $l_1=2l_0-l_2=3$ and $p_{\max}=\min(2\abs{l_0},\abs{l_2})=1\,$. Furthermore, we define the modes as $\mathcal A = LG_{0, 1}(w_2=w)$, $\mathcal B_0 = LG_{0, 3}(w_1)$, $\mathcal B_1 = LG_{1, 3}(w_1)$ and $\mathcal C = LG_{2, 3}(w_1)\,$. Thus $\mathcal C$ is a radial mode outside the first-order support for coupling to $\mathcal A$. The short-distance calculation certifies the pairs $\mathcal A|\mathcal B_0$ and $\mathcal A|\mathcal B_1$, while it is inconclusive for $\mathcal A|\mathcal C$. By Sec.~\ref{sec:pairwise_partitions}, these two pairs already certify six of the seven bipartitions; the remaining cut is $\mathcal A\mathcal B_0\mathcal B_1|\mathcal C$. In Fig.~\ref{fig:fully_wit} we apply FullyWit to the reduced covariance matrix of these four modes and optimize a witness separately for each bipartition. The numerical curves form three groups:

\begin{itemize}
    \item $\mathcal A\mathcal B_1\mathcal C|\mathcal B_0$, $\mathcal A\mathcal B_1|\mathcal B_0\mathcal C$, $\mathcal A\mathcal C|\mathcal B_0\mathcal B_1$ and $\mathcal A|\mathcal B_0\mathcal B_1\mathcal C$ show the lowest witness values and hence the largest margins below the separability threshold. These share the fact that $\mathcal A$ and $\mathcal B_0$ belong to different parties, giving a stronger second-moment entanglement certification across these partitions.
    \item $\mathcal A\mathcal B_0\mathcal C|\mathcal B_1$ and $\mathcal A\mathcal B_0|\mathcal B_1\mathcal C$ show intermediate witness values, still maintaining a linear start. Here the entangled pair $\mathcal A|\mathcal B_1$ crosses the cut, while $\mathcal A$ and $\mathcal B_0$ belong to the same party.
    \item $\mathcal A\mathcal B_0\mathcal B_1|\mathcal C$ shows the smallest margin below the separability threshold, with a numerical onset consistent with a nonlinear fall. Here both first-order entangled pairs lie within $\mathcal A\mathcal B_0\mathcal B_1$, so the individual-pair argument gives no first-order certificate across this cut.
\end{itemize}

The crossing-pair argument explains why the first six cuts admit linear violations, but does not predict the optimized witness values. The small violation for $\mathcal A\mathcal B_0\mathcal B_1|\mathcal C$ illustrates that modes outside the first-order radial support can nevertheless participate in entanglement at finite distance.

\subsection{Genuine multipartite entanglement of selected modes}

Finally, we apply the biseparable-class optimization of Sec.~\ref{sec:covariance_witnesses} to the reduced selected modes. Unlike testing each cut separately, a MultiWit violation excludes all mixtures in Eq.~\eqref{eq:biseparable_mixture} and therefore certifies GME after the remaining field has been discarded.

In Fig.~\ref{fig:multi_wit}, we show the witness for sets of the form $\{LG_{0, 1}(w_2 = w)\} \cup \{LG_{p, 3}(w_1) \ \vert \ p \le P\}$, with $P$ varying from $0$ to $3$, while maintaining $l_0 = 2$. The $P=0$ case is bipartite; for $P\ge1$, a violation certifies genuine multipartite entanglement. The numerical results display a pattern related to the first-order radial support: the sets with $P \le p_{\text{max}} = 1$ have lower witness values and hence larger margins below the separability threshold.

In particular, for $P=1$ Eq.~\eqref{eq:main_grouped_parties} certifies the pure-state grouping $\mathcal A|\mathcal B_0|(\mathcal B_1\mathcal R)$, whereas a MultiWit violation certifies the reduced three-mode state on $\mathcal A|\mathcal B_0|\mathcal B_1$. The latter shows that genuine multipartite entanglement survives discarding the remaining field modes. Violations for larger selected sets, when present, involve modes outside the first-order support and are finite-distance results within the Bogoliubov approximation.

\section{Experimental proposal}
\label{sec:experimental_implementation}

\begin{figure*}[!t]
    \centering
    \includegraphics[width=\linewidth]{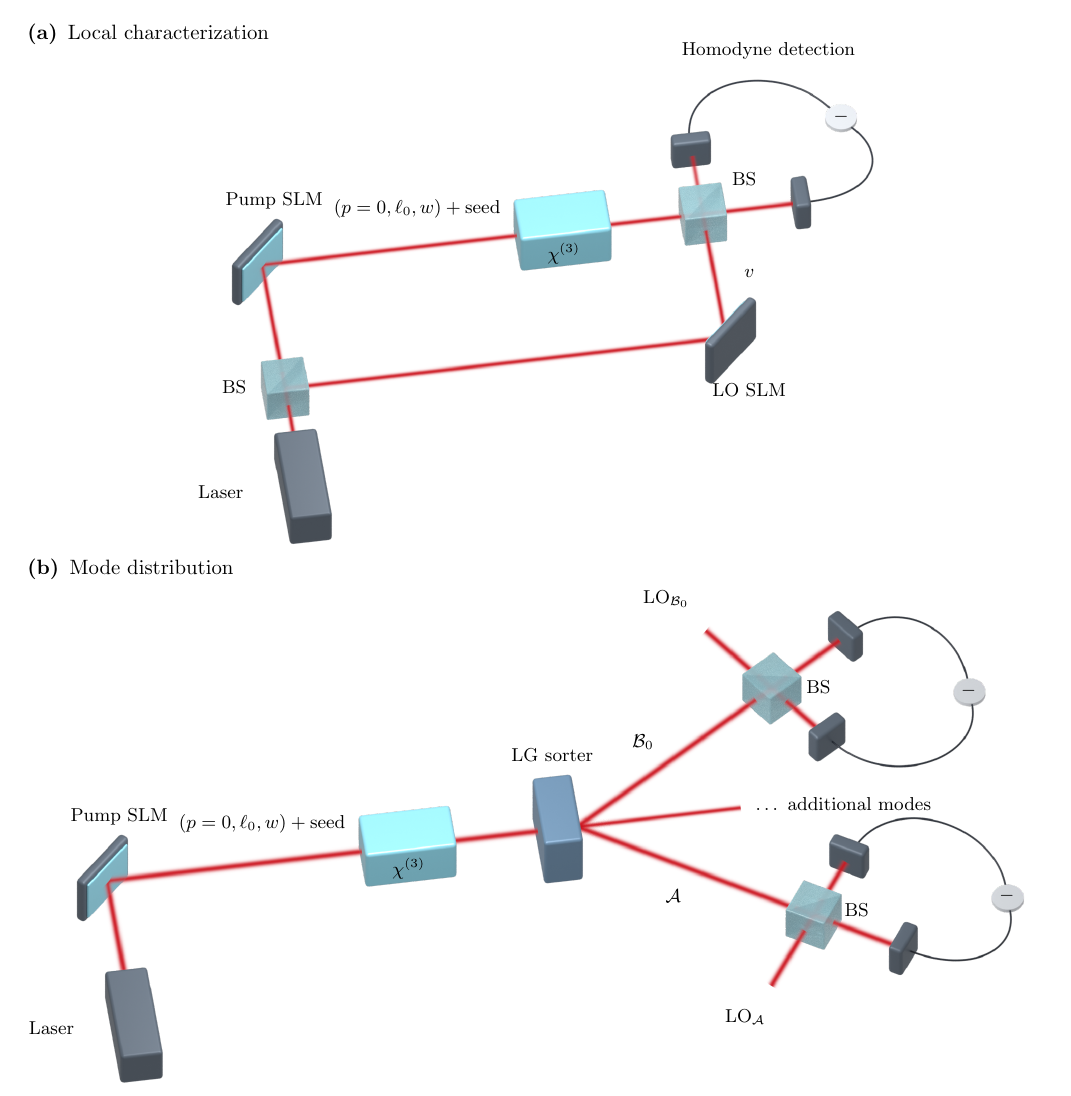}
    \caption{Experimental implementation schemes. (a) Local characterization by balanced homodyne detection with a spatially shaped local oscillator in mode $v$. (b) Mode distribution using a Laguerre--Gaussian (LG) mode sorter, followed by balanced homodyne detection of the separated modes with their respective local oscillators. In both schemes, a spatial light modulator (SLM) shapes the pump before the $\chi^{(3)}$ nonlinear medium. BS denotes a beam splitter, and LO denotes a local oscillator.}
    \label{fig:experimental_implementation}
\end{figure*}

Figure~\ref{fig:experimental_implementation} outlines two complementary implementations of the proposed entanglement scheme: local characterization of the covariance matrix and spatial distribution of the correlated modes. In both cases, a spatial light modulator (SLM) prepares an LG pump that propagates through a $\chi^{(3)}$ nonlinear medium. The interaction generates correlations between transverse modes, which are subsequently probed by balanced homodyne detection with a spatially shaped local oscillator (LO).

For this preparation stage, the required interaction strength can be assessed through the accumulated nonlinear phase. Fontaine et al.~\cite{Fontaine2020Bogoliubov} investigated a propagating fluid of light in a hot rubidium vapor cell of length $L=7.5\,\mathrm{cm}$ at a wavelength $\lambda=780\,\mathrm{nm}$, finding a nonlinear index change proportional to the intensity, consistent with an unsaturated Kerr response. Taking $|\Delta n|\simeq5\times10^{-6}$ as a representative value within their experimental range gives the effective phase estimate
\begin{equation}
    \Phi_{\mathrm{NL}}\simeq\frac{2\pi}{\lambda}L|\Delta n|
    \simeq 3.0\,\mathrm{rad}.
\end{equation}
In our diffraction-free model, the local phase magnitude is $|Z||u(\mathbf r)|^2$. For an $LG_{0,l}$ pump, its peak value is
\begin{equation}
    \Phi_{\mathrm{pk}}=|Z|\frac{2}{\pi}
    \frac{|l|^{|l|}e^{-|l|}}{|l|!},
\end{equation}
where the last fraction is unity for $l=0$. At $|Z|=1$, the peak phases are approximately $0.64$, $0.23$, and $0.17\,\mathrm{rad}$ for $|l|=0$, $1$, and $2$, respectively. Their smallness relative to the experimental phase scale supports the accessibility of $|Z|=1$, where our finite-distance calculations already predict clear entanglement signatures.

Transferring this estimate to the proposed setup requires accounting for the power delivered to the medium after LG preparation and filtering, as well as the change in peak intensity associated with the selected spatial profile. Preparation losses can in principle be compensated by a more powerful input laser. The comparison therefore provides an estimate of a feasible interaction strength, while the observable entanglement also depends on propagation and detection losses and noise from the medium.

The same experiment also provides a transverse scale for estimating diffraction. At the cell entrance, Fontaine et al.~\cite{Fontaine2020Bogoliubov} report an elliptical beam with $1/e^2$ intensity radii of $500\,\mu\mathrm{m}$ and $1\,\mathrm{cm}$. Taking the smaller radius as an illustrative waist $w$ for our circular LG pump gives a matched partner waist $w_1=w/\sqrt{3}\simeq 289\,\mu\mathrm{m}$. At $780\,\mathrm{nm}$, the corresponding Rayleigh ranges are $z_R=\pi w^2/\lambda\simeq 1.0\,\mathrm{m}$ and $z_{R,1}=z_R/3\simeq 0.34\,\mathrm{m}$. Since the reference mode $\mathcal A$ has waist $w_2=w$, the cell spans about $0.075\,z_R$ for both the pump and $\mathcal A$ and about $0.22\,z_{R,1}$ for the radial partner modes. These ratios characterize diffraction on the waist scales; finer structure generated by the nonlinear phase can diffract over shorter distances, as discussed in Appendix~\ref{app:no_diffraction}.

To characterize the resulting state locally, we propose the arrangement in Fig.~\ref{fig:experimental_implementation}(a), which uses a Mach--Zehnder interferometer. One arm contains the pump preparation and nonlinear medium, while the other supplies the LO, shaped by a second SLM into a configurable mode $v$. Balanced homodyne detection measures $\Gamma(v,v)$ (see Eq.~\eqref{eq: covariance}), giving the diagonal covariance elements. The off-diagonal elements follow from
\begin{equation}
\begin{aligned}
        \Gamma(v_1,v_2) = &\Gamma\left(\frac{v_1+v_2}{\sqrt{2}},\frac{v_1+v_2}{\sqrt{2}}\right) \\&- \frac{1}{2}\left[ \Gamma(v_1,v_1) + \Gamma(v_2,v_2) \right],
\end{aligned}
\end{equation}
which requires an additional measurement with the LO prepared in the superposition $(v_1+v_2)/\sqrt{2}$. These measurements provide the covariance information needed to evaluate the entanglement criteria discussed above.

Coherent seeds collinear with the pump can stimulate the process by populating the selected modes $\mathcal{A},\mathcal{B}_0,\dots$, facilitating the experimental detection. Appendix~\ref{app:coherent_seed} examines this possibility and shows that, for the configurations considered, a seed fraction of $1\%$ of the total input power produces only modest changes in the finite-distance witness curves.

Beyond local characterization, the correlated modes can be distributed to separate parties using the arrangement in Fig.~\ref{fig:experimental_implementation}(b). An LG mode sorter~\cite{fontaine_laguerre_gaussian_2019,bearne_diffractive_2026,mirhosseini_efficient_2013,berkhout_efficient_2010,jia_vector_2023} spatially separates the output components, allowing each party to perform balanced homodyne detection with an appropriate LO. The two arrangements thus address both the covariance measurements needed to certify the generated entanglement and the spatial separation needed to make its modes individually accessible for potential non-local quantum information protocols.

\section{Conclusion}
\label{sec:conclusion}

We have derived a chiral selection rule for the first-order anomalous correlations generated by self-phase modulation of a coherent OAM beam. At specified detection waists, radial--angular coupling fixes a finite set of partner modes with nonzero linear Duan violations. These violations guarantee pairwise entanglement and full inseparability of the selected set at sufficiently short nonzero distances. For the full pure field, assigning the remaining modes to one selected party also gives a multipartite grouping whose entanglement follows from the same correlations.

Finite-distance calculations within the diffraction-free Bogoliubov approximation show that the first-order support remains useful for organizing entanglement certification. In the configurations investigated, optimized witnesses reveal weaker violations involving modes outside that support, so it is not an exact finite-distance entanglement boundary. Witnesses valid for all biseparable states additionally certify GME in reduced sets of detection modes. These conclusions concern the certification margins of the chosen criteria and do not quantify the amount of entanglement.

The results identify OAM and the detection profiles as useful parameters for controlling the radial mode combinations that admit covariance-based entanglement certification. They can be investigated in four-wave mixing and self-phase modulation settings where diffraction over the interaction region and fluctuation backreaction are sufficiently small. These findings can be useful for quantum information protocols that require control over multipartite entanglement.

\appendix

\section{Diffraction-free Bogoliubov evolution}
\label{app:no_diffraction}

The first-order calculation in the main text retains diffraction and all nonlinear fluctuation terms in Eq.~\eqref{eq:bg_full_fluctuation}. Here we start from the same full fluctuation equation and adopt a different approximation: we neglect diffraction over the interaction region and discard the quadratic and cubic fluctuation terms $\Omega$ of Eq.~\eqref{eq:bg_remainder}, while retaining the full dependence on propagation distance. The resulting Bogoliubov equations are solved without expanding in the propagation distance or the accumulated classical nonlinear phase.

With the same decomposition $\mathcal E=\alpha_c+\eta$ and input $\alpha(\mathbf r)=\alpha_c(\mathbf r,0)$, these equations are
\begin{equation}
\begin{aligned}
    2ik\partial_z\alpha_c&=g|\alpha_c|^2\alpha_c,\\
    2ik\partial_z\eta&=g\left(2|\alpha_c|^2\eta+\alpha_c^2\eta^\dagger\right).
\end{aligned}
    \label{eq:nodiff_bog_equations}
\end{equation}
The classical intensity is constant along propagation. Defining
\begin{equation}
    I(\mathbf r)=|\alpha(\mathbf r)|^2,
    \qquad \theta(\mathbf r,z)=\frac{gz}{2k}I(\mathbf r),
\end{equation}
we have
\begin{equation}
    \alpha_c(\mathbf r,z)=\alpha(\mathbf r)e^{-i\theta(\mathbf r,z)}.
    \label{eq:nodiff_classical}
\end{equation}

\subsection{Local fluctuation solution}

Each transverse position evolves independently. Introduce the rotating-frame fluctuation $b(\mathbf r,z)=e^{i\theta(\mathbf r,z)}\eta(\mathbf r,z)$. Suppressing the position argument temporarily, Eq.~\eqref{eq:nodiff_bog_equations} becomes
\begin{equation}
    2ik\partial_z b=g\left(Ib+\alpha^2b^\dagger\right).
\end{equation}
Together with its conjugate, this gives
\begin{equation}
    2ik\partial_z\begin{pmatrix}b\\b^\dagger\end{pmatrix}
       =gM\begin{pmatrix}b\\b^\dagger\end{pmatrix},
    \qquad
    M=\begin{pmatrix}I&\alpha^2\\-\alpha^{*2}&-I\end{pmatrix}.
\end{equation}
The matrix is independent of $z$ and satisfies $M^2=0$. Therefore its exponential terminates exactly:
\begin{equation}
    \exp\left(-\frac{igz}{2k}M\right)
       =\mathbb I-\frac{igz}{2k}M.
\end{equation}
Returning to $\eta$ and writing $\eta_0(\mathbf r)=\eta(\mathbf r,0)$, we obtain
\begin{equation}
\begin{aligned}
    \eta(\mathbf r,z)=e^{-i\theta}\bigg[&(1-i\theta)\eta_0(\mathbf r)\\
       &-i\frac{gz}{2k}\alpha^2(\mathbf r)\eta_0^\dagger(\mathbf r)\bigg].
\end{aligned}
    \label{eq:nodiff_fluctuation_solution}
\end{equation}
Although the rotating-frame coefficients are linear in $z$, this is not a first-order Taylor approximation. It is the exact solution of the diffraction-free linearized fluctuation equations.

Writing $\eta=A\eta_0+B\eta_0^\dagger$, the coefficients are
\begin{equation}
    A=e^{-i\theta}(1-i\theta),\qquad
    B=-i\frac{gz}{2k}\alpha^2e^{-i\theta}.
\end{equation}
They obey $|A|^2-|B|^2=1$, so the field commutator is preserved. For the coherent input, $\langle\eta_0\rangle=0$ and hence $\langle\eta(z)\rangle=0$. Thus the mean field equals $\alpha_c$ within this approximation.

\subsection{Field and mode correlations}

The initial fluctuations satisfy
\begin{equation}
\begin{aligned}
    \langle\eta_0(\mathbf r_1)\eta_0^\dagger(\mathbf r_2)\rangle
       &=\delta^{(2)}(\mathbf r_1-\mathbf r_2),\\
    \langle\eta_0^\dagger(\mathbf r_2)\eta_0(\mathbf r_1)\rangle
       &=\langle\eta_0(\mathbf r_1)\eta_0(\mathbf r_2)\rangle=0.
\end{aligned}
\end{equation}
Consequently the correlations are
\begin{equation}
\begin{aligned}
    F(\mathbf r_1,\mathbf r_2;z)
       &=A(\mathbf r_1)B(\mathbf r_2)\delta^{(2)}(\mathbf r_1-\mathbf r_2),\\
    H(\mathbf r_1,\mathbf r_2;z)
       &=B(\mathbf r_1)B^*(\mathbf r_2)\delta^{(2)}(\mathbf r_1-\mathbf r_2).
\end{aligned}
\end{equation}
Substituting the coefficients gives
\begin{equation}
\begin{aligned}
    F(\mathbf r_1,\mathbf r_2;z)
       ={}&\left[-\frac{igz}{2k}-\left(\frac{gz}{2k}\right)^2I(\mathbf r_1)\right]\\
       &\times\alpha_c^2(\mathbf r_1,z)\delta^{(2)}(\mathbf r_1-\mathbf r_2),\\
    H(\mathbf r_1,\mathbf r_2;z)
       ={}&\left(\frac{gz}{2k}\right)^2I^2(\mathbf r_1)
                         \delta^{(2)}(\mathbf r_1-\mathbf r_2).
\end{aligned}
    \label{eq:nodiff_field_correlations}
\end{equation}
Using the main-text scaling, $\theta=Z|u|^2$ and $u_Z=u\exp(-iZ|u|^2)$, projection onto the detection modes yields
\begin{equation}
\begin{aligned}
    f(v_1,v_2)={}&-iZ(v_1,v_2;u_Z,u_Z)\\
                &-Z^2(v_1,v_2,u_Z;u_Z,u_Z,u_Z),\\
    h(v_1,v_2)={}&Z^2(v_1,u_Z,u_Z;v_2,u_Z,u_Z).
\end{aligned}
    \label{eq:nodiff_mode_correlations}
\end{equation}
The phases in $h$ cancel, whereas those in $f$ retain the nonlinear classical evolution. Expanding Eq.~\eqref{eq:nodiff_mode_correlations} at small $Z$ recovers $f=-iZ(v_1,v_2;u,u)+\mathcal{O}(Z^2)$ and $h=\mathcal{O}(Z^2)$, in agreement with the first-order calculation in the main text. At finite distance, these expressions are exact only within the diffraction-free Bogoliubov approximation, not for the full quantum Kerr equation.

The approximation requires negligible diffraction of the relevant transverse structures over the interaction length, as well as sufficiently small fluctuation interactions and backreaction on the classical field. A large pump Rayleigh range is a useful initial scale, but finer structure generated by the nonlinear phase can have shorter diffraction lengths. The derivation applies to any coherent input profile, including a coherent pump plus seeds.

The Bogoliubov approximation also permits retaining diffraction in both the classical background and the linear fluctuation equations. Quantum correlations can then be computed numerically by propagating the corresponding Green functions~\cite{PhysRevA.62.033816, Eric_Lantz_2004}, or by back-propagating the detection functions with the adjoint fluctuation equations~\cite{PhysRevA.51.817, Ray_Kuang_Lee_2004} and evaluating their correlations in the known input state. Such a numerical treatment with diffraction lies beyond the scope of the present work, which uses the analytical diffraction-free solution above.

\section{Coherent seeding of the selected modes}
\label{app:coherent_seed}

\begin{figure*}[!t]
    \centering
    \includegraphics[width=\linewidth]{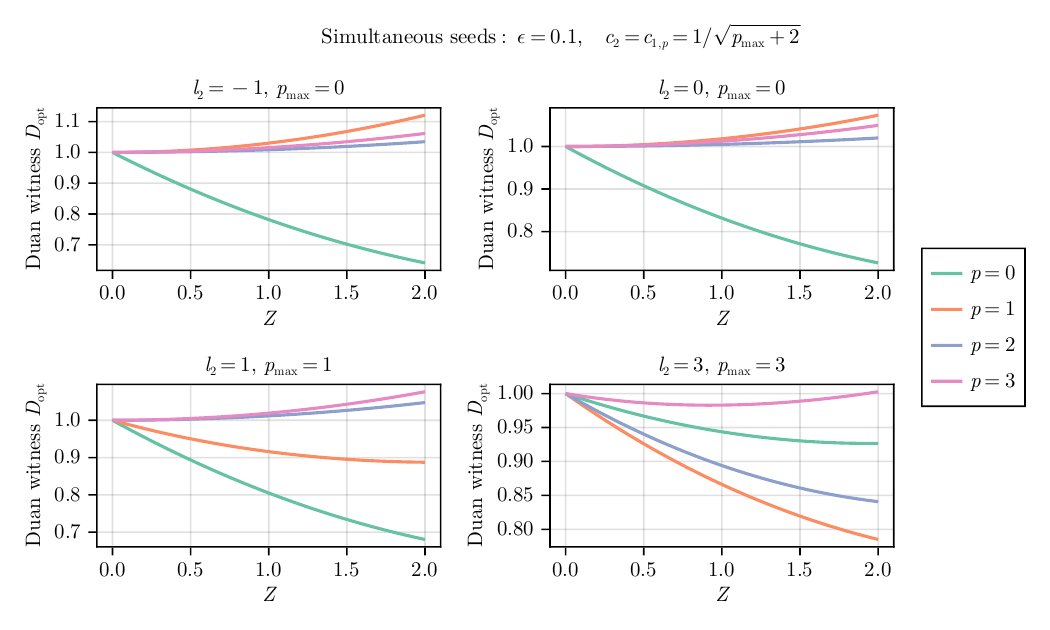}
    \caption{Phase-optimized Duan witness with simultaneous coherent seeding of the reference mode and all radial partners $p=0,\ldots,p_{\max}$, with seed amplitude $\epsilon=0.1$ (power fraction $\epsilon^2=0.01$) and equal amplitudes and phases as in Eq.~\eqref{eq:multiseed_input}. The pump OAM is $l_0=2$, and the detection modes and waists match Fig.~\ref{fig:duan}. All curves within each panel use the same seeded input. Values below one certify two-mode entanglement within the diffraction-free Bogoliubov model.}
    \label{fig:duan_seeded}
\end{figure*}

Coherent seeds in all selected detection modes stimulate the investigated process, which may assist the experimental implementation discussed in Sec.~\ref{sec:experimental_implementation}. An unseeded pump with OAM $l_0$ has no coherent mean in the detected OAM families $l_1+l_2=2l_0$, with $l_2\ne l_0$, although these modes can be correlated and measured with an external local oscillator. Here we add weak coherent seeds directly to the reference mode and every radial partner selected by the first-order rule, and assess their effect on the entanglement witnesses.

We retain the mode definitions of Sec.~\ref{sec:selection rules}: the pump profile is $LG_{0,l_0}(\mathbf r;w)$, the reference mode is $v_2=LG_{0,l_2}(\mathbf r;w_2)$, and a radial partner is $v_1=LG_{p,l_1}(\mathbf r;w_1)$, with
\begin{equation}
    l_1=2l_0-l_2,\qquad
    \frac{1}{w_1^2}=\frac{2}{w^2}+\frac{1}{w_2^2}.
    \label{eq:seed_detection_family}
\end{equation}
We seed the reference mode $\mathcal A$ and all radial partners $\mathcal B_p$, $p=0,\ldots,p_{\max}$, with the cutoff defined in Eq.~\eqref{eq:radial_cutoff}.
Choosing equal amplitudes and phases gives the normalized seed profile
\begin{equation}
    s=\frac{v_2+\sum_{p=0}^{p_{\max}}LG_{p,l_1}(\mathbf r;w_1)}{\sqrt{N_s}},
    \qquad N_s=p_{\max}+2.
    \label{eq:multiseed_input}
\end{equation}
The seed modes are mutually orthogonal and orthogonal to the pump. For mutually coherent, collinear fields with the same frequency, polarization, and temporal mode, we therefore write
\begin{equation}
    \alpha(\mathbf r)=\norm{\alpha}
    \left[\sqrt{1-\epsilon^2}\,LG_{0,l_0}(\mathbf r;w)+\epsilon\,s(\mathbf r)\right],
    \label{eq:seed_input}
\end{equation}
where $0\le\epsilon<1$ is the seed amplitude parameter and $\epsilon^2$ is the fraction of the total input power assigned to the seed. Each seeded mode then has the input mean amplitude $\norm{\alpha}\epsilon/\sqrt{N_s}$. We hold the total occupation $\norm{\alpha}^2$ fixed, so the comparison with the unseeded case uses the same propagation coordinate,
\begin{equation}
    u=w\left[\sqrt{1-\epsilon^2}\,LG_{0,l_0}(\mathbf r;w)+\epsilon\,s\right],
    \qquad Z=\frac{gz\norm{\alpha}^2}{2kw^2}.
    \label{eq:seed_normalization}
\end{equation}

The leading change in the selected correlations follows directly from their nonlinear overlap. Since neither seed family has OAM $l_0$, the pump--seed terms in $u^2$ cannot contribute to the channel $l_1+l_2=2l_0$. The two seed families do have complementary OAM, however, and their product contributes at order $\epsilon^2$. For a detected partner $v_1=LG_{p,l_1}(\mathbf r;w_1)$, let $C_p$ denote the unseeded coefficient defined in the main text. Its seeded counterpart is
\begin{equation}
    \begin{aligned}
        C_p(\epsilon)&=(v_1,v_2;u,u)\\
        &=(1-\epsilon^2)C_p+
                    \frac{2\epsilon^2}{N_s}\sum_{p'=0}^{p_{\max}}K_{pp'},\\
        K_{pp'}&=w^2\bigl(v_1,v_2;v_2,LG_{p',l_1}(w_1)\bigr).
    \end{aligned}
    \label{eq:seed_overlap}
\end{equation}
Here $p'$ labels the seeded radial partners, and $K_{pp'}$ uses the spatial overlap notation of the main text, with spatial arguments suppressed.
The first-order anomalous correlation and phase-optimized Duan witness are consequently
\begin{equation}
    \begin{aligned}
        f(v_1,v_2)&=-iZC_p(\epsilon)+\mathcal{O}(Z^2),\\
        D_{\mathrm{opt}}&=1-2|Z|\,|C_p(\epsilon)|+\mathcal{O}(Z^2).
    \end{aligned}
    \label{eq:multiseed_correlations}
\end{equation}
Thus the selected overlap coefficients change by $\mathcal{O}(\epsilon^2)$, with no correction linear in $\epsilon$. These changes arise because the seed participates in the nonlinear interaction; a coherent displacement alone does not change centered correlations. The weighted overlaps $K_{pp'}$ need not vanish beyond the original radial cutoff, so weak seeding preserves the dominant correlation pattern without preserving every first-order zero exactly. Small absolute corrections can also matter for pairs close to a witness threshold.

To assess the effect at finite distance, we insert Eq.~\eqref{eq:seed_normalization} into the diffraction-free Bogoliubov correlation formulas of Appendix~\ref{app:no_diffraction}. Figure~\ref{fig:duan_seeded} uses the same pump and detection parameters as Fig.~\ref{fig:duan}, with $w_2=w$, $w_1=w/\sqrt{3}$, and $\epsilon=0.1$: one percent of the total input power is shared equally among all $N_s$ selected modes. Every curve within a panel describes this same seeded input state, including curves for unseeded radial partners outside the first-order support.

Compared with the unseeded results in Fig.~\ref{fig:duan}, weak seeding leaves the overall certification pattern largely unchanged. Pairs with substantial unseeded violations retain clear violations, and the relative behavior of the radial curves remains similar. The small shifts can nevertheless affect pairs close to the separability threshold. For example, the $l_2=3$, $p=3$ pair ceases to violate the witness near the end of the plotted interval, although it remains certified at shorter distances. Failure to violate the witness there is inconclusive about entanglement.

These examples show that direct coherent references can be introduced simultaneously into all modes selected by the first-order rule with only modest changes to the finite-distance witness curves for the seed strength and phases considered. This agrees qualitatively with the weak-seed analysis above, while the near-threshold changes emphasize that preserving the dominant pattern does not imply an exact finite-distance selection rule.

\bibliography{refs}

\end{document}